# Detection of Molecular Scattering Field by Polarization Analysis of Sky Background During the Transitive Twilight and Temperature Measurements near the Stratopause

Ugolnikov O.S.[1], Maslov I.A.[1,2], Kozelov B.V.[3]

[1]Space Research Institute, Russian Academy of Sciences
Profsoyuznaya st., 84/32, Moscow 117997 Russia
[2]Moscow State University, Sternberg Astronomical Institute,
Universitetsky prosp., 13, Moscow 119234 Russia
[3]Polar Geophysical Institute,
Akademgorodok st., 26a, Apatity, Murmansk distr. 184209 Russia

E-mail: ougolnikov@gmail.com, imaslov@iki.rssi.ru, boris.kozelov@gmail.com

The simplest version of the method of detection of single molecular scattering basing on the polarization measurements of the twilight sky background by all-sky cameras is considered. The method can be used during the transitive twilight (solar zenith angles 94°-98°), when effective single scattering takes place in upper stratosphere and lower mesosphere. The application of the method to the multi-year measurements in the vicinity of Moscow and Apatity allows determination of temperature of these atmospheric layers and estimation of contribution and properties of multiple scattering during the transitive twilight period.

1. INTRODUCTION

Study of middle and upper atmospheric layers becomes more actual in a present time. First, modern techniques allow doing it with high accuracy. Second, climate changes including the anthropogenic ones influence the upper layers. As an example, the increase of carbon dioxide contributes to the greenhouse effect near the Earth's surface but causes the cooling of upper [1] and middle atmosphere [2]. Negative temperature trends were fixed in upper stratosphere [3, 4] and mesosphere [5-7].

Temperature analysis of stratosphere is especially interesting in polar regions due to events of "sudden stratosphere warming" [8-10] when the temperature can increase by several dozens of degrees. In several cases the effect can reach 50 degrees [11]. This is accompanied by rapid changes of dynamical structure of stratosphere [7, 12].

Stratopause and mesosphere are less available for direct studies than lower and middle stratosphere. The basic information about physical condition of these layers is provided by optical remote sounding. The measurements are hold by lidars [10, 11, 13, 14] and satellites. Global temperature data is now provided by SABER (*Sounding of the Atmosphere Using Broadband Emission Radiometry*) instrument onboard TIMED (*Thermosphere, Ionosphere, Mesosphere, Energetics and Dynamics*) satellite [15] and MLS (*Microwave Limb Sounder*) instrument onboard EOS (*Earth Observing System*) Aura satellite [16]. In both cases the temperature is defined by microwave sounding at certain altitude above the Earth's limb.

Temperature can be also measured by Rayleigh lidar [13, 17], the vertical profile of molecular scattering is built for this purpose. The same method can be used for measurements of scattered solar radiation after the sunset, as the altitude of Earth's shadow increases. Twilight analysis is more efficient if the most part of sky hemisphere is observed and the polarization is fixed along with intensity of scattered emission.

The basic problem of twilight method of atmosphere sounding is significant contribution of multiple scattering to the twilight background. Necessity of its account was noticed in XX century [18] but its value was underestimated for a long time. Polarization study [19, 20] had shown that multiple scattering contribution is high even at sunset, reaching 50% in blue part of spectrum. It is the multiple scattering that restricts the altitude applicability range of twilight sounding method: single scattering vanishes on the multiple scattered background when the effective altitude of single scattering reaches 90 km. In the zenith point this corresponds the solar depression under the horizon at 9° [21]. This moment falls to the short period (solar zenith angles from 98° to 100°), when the single scattering is observed in the dusk area but disappears in the opposite sky part. For this period the multiple scattering separation procedure was established; it allows retrieving the temperature profile in upper mesosphere, from 70 to 85 km [22, 23].

During the lighter twilight stage (solar zenith angles 94°-98°) single scattering contributes to the total sky background over the most part of the sky, the dependence on the sky point is complicated. The basic aim of this work is to build the simple method of single and multiple scattering separation during this period corresponding to the altitudes of upper stratosphere and lower mesosphere. The accuracy criteria can be the similarity of dependence of single scattering intensity on the Earth's shadow altitude and the pressure profile and also adequate Boltzmann temperature values. The method is applied to the results of polarization measurements of the sky background using all-sky cameras.

2. OBSERVATIONS

The paper is experimentally based on two all-sky cameras designed for polarization measurements. First camera is installed in Moscow region (Chepelevo, 55.2°N, 37.5°E) and works since 2011. The data obtained by this camera were used for retrieval of temperature profiles in upper mesosphere [21, 22, 24]. The field diameter is about 140°, the instrumental band has the effective wavelength 540 nm and FWHM 90 nm. During the spring and early summer of 2016 the camera worked with three-color (RGB) CCD-detector in order to search for effects of stratospheric aerosol scattering at the altitudes up to 35 km [25]. In this paper we use the data in G band of this detector that is close to the only instrumental band of this camera during other observational periods (until 2015 and since mid-summer of 2016).

Second camera is installed in Apatity (67.6°N, 33.4°E) in early 2015. It was also used for temperature measurements in mesosphere [23]. Its characteristics are similar to Moscow camera, effective wavelength is equal to 530 nm. The field diameter is 180°, the intensity and polarization measurements at zenith distances up to 60° are used in this work. During the epochs of bright noctilucent clouds the data of both cameras were used for polarization study of microphysical properties of cloud particles [26], the results were confirmed later by noctilucent clouds color analysis [27].

Cameras optical design is described in papers cited above. Both devices consist of three consecutive wide-field lenses. The first one collects the emission of sky hemisphere, the second lens creates its remote image, reducing the axial angle of light rays down to 5°-10°. This makes possible to install rotating polarization filter beyond the second lens. Finally, third lens creates the sky image on CCD-detector. The image diameter varies a little for different cameras and detectors, being close to 500 pixels in average. For study of the twilight sky without short-scale structures like noctilucent clouds the measurement data with resolution 5° is used.

Variation of exposure time allows holding the measurements during the whole twilight period, the beginning and the end of a day and the night. Analysis of star images in nighttime frames fixes the camera position and field curvature. Star photometry is used to find the product of camera flat field and atmospheric transparency at different point zenith angles, that is necessary for the procedure described here. As in [25], the measurements in solar vertical are used in this paper. In this case the sky position is characterized by zenith angle ζ, that is positive in dusk region and negative in opposite part of the vertical.

Figure 1 shows the dependencies of sky polarization on solar zenith angle for different solar vertical points for the evening twilight of May, 31, 2014, the vicinity of Moscow. The evolution of this value was described in previous papers [19, 20, 28, 29]. It reflects the influence of two factors reducing the sky polarization: scattering on aerosol particles and multiple scattering in atmosphere. Aerosol scattering reveals itself during the day and very light twilight, it causes the decrease of polarization near the zenith in this period (solar zenith angle up to 91°-92° [30]). This time Sun illuminates the troposphere containing the most part of aerosol particles. Scattering on stratospheric aerosol is noticeable during the darker stage of twilight, at solar zenith angles 92°-95° [29], this effect increases in red spectral region [25].

During even deeper twilight, when effective altitude of single scattering exceeds 40 km, the aerosol depolarization effects fade away. Fast decrease of polarization observed in this period was incorrectly related with aerosol before, it is caused by increase of multiple scattering contribution. In classification [28] this period is called transitive twilight. This time the effective scattering layer moves from stratosphere to mesosphere. In this work we analyze the observational data of this twilight stage.

When solar zenith angle reaches 99°, the fall of sky polarization gets slower (or stops for a little time if the night sky background is weak). This corresponds to the dark stage of twilight when single scattering vanishes on the less polarized multiple scattering background. Further polarization decrease is related with raising contribution of night sky background.

All twilight stages shift depending on sky point position (see the curves for ζ=±45° in Figure 1). Effective altitude is less in dusk region and all twilight stages come later there. During the transitive twilight considered here, single scattering contribution in dusk region is significantly higher than in opposite sky part, that is used to build the method described in the following chapter.

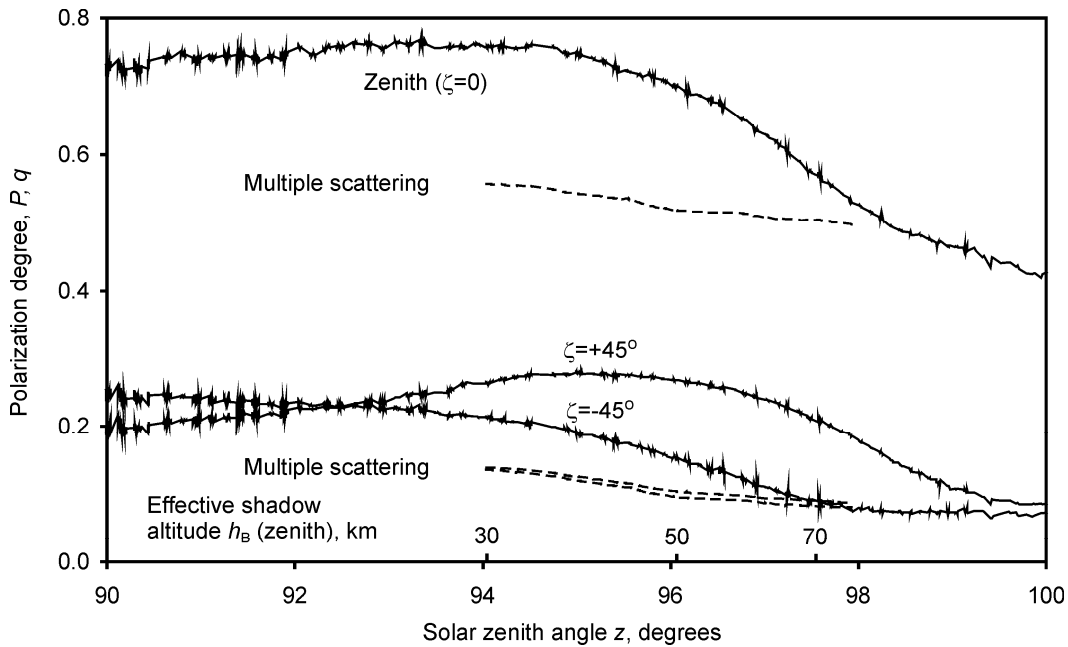

*Figure 1. Polarization of total background and its multiple scattering in three solar vertical points during evening twilight of May, 31, 2014, vicinity of Moscow.*

## 3. SINGLE SCATTERING DETECTION METHOD

The problem of separation of different components of the twilight glow (single molecular scattering, multiple scattering, aerosol scattering for some altitude ranges) is the most complicated problem of twilight sounding method. Results of numerical modeling of multiple scattering [31, 32] strongly depend on input parameters, those are generally unknown. Moreover, the models are applicable until the beginning of transitive twilight (solar zenith angle 96°), the accuracy rapidly falls after that due to fast increase of effective scattering order. The single scattering layer just crosses the stratopause this time.

The methods applying some assumptions for the multiple scattering properties during the transitive and dark twilight are more effective for optical study of mesosphere. One simplification meant the equality of derivatives of intensity logarithm of multiple scattering ln *j* on the solar zenith angle *z* in opposite solar vertical points. It actually means the proportionality of intensity values *j*(ζ, *z*) and *j*(–ζ, *z*) in these points for a short range of *z*. These models were suggested in XX century [33] and used for analysis of all-sky cameras data at solar zenith angles more than 97°, i.e. for upper mesosphere research [21, 22]. During the lighter twilight stages such models are less reliable due to contribution of tropospheric aerosol to the multiple scattering.

Another property of multiple scattering detected experimentally [19] and confirmed in [21, 22] is the polarization symmetry of multiple scattering in solar vertical relatively the zenith. This property is the base of separation method suggested here and primal estimation of Boltzmann temperature in stratosphere and mesosphere, that is to be clarified later.

Let us consider the simple model of single molecular scattering used at the first stage of procedure (Figure 2). We assume that the atmosphere of the Earth is transparent for tangent solar emission transferred above the definite altitude $h_P$ and totally absorbs the emission below it. This altitude can be considered as constant during all twilight stages except the light twilight. Following [22], we assume it to be equal to 14 km. This case the intensity of single molecular scattering $J(\zeta, z)$ in solar vertical point with coordinate ζ at solar zenith angle *z* (after the correction by flat field and atmosphere extinction above the observer) is proportional to the air column density above the shadow (or twilight layer baseline) altitude $h_B(\zeta, z)$ or the pressure at this altitude:

$$J(\zeta, z) = const \frac{p(h_B(\zeta, z))(1.06 + \cos^2\theta(\zeta, z))}{\cos\zeta}. \tag{1}$$

The last term in the numerator is related to the Rayleigh scattering function of two-atomic air molecules. Scattering angle θ is equal to

$$\theta = z - \zeta - \rho. \tag{2}$$

Here ρ is the refraction angle of tangent solar emission. For perigee altitudes $h_P$ corresponding to the stratosphere this angle does not exceed 0.2° and can be neglected. Substitution of (2) into (1) gives the intensity profile of single scattering along the solar vertical in definite twilight moment:

$$J(\zeta, z) = J(0, z) \cdot \frac{p(h_B(\zeta, z)) \cdot (1.06 + \cos^2(z - \zeta))}{p(h_B(0, z)) \cdot \cos\zeta \cdot (1.06 + \cos^2 z)}. \tag{3}$$

This formula can be used for point zenith distances ζ up to ±70°. We don't take into account the atmospheric extinction after the scattering event since it is corrected by the star image analysis in the nighttime images. The shadow altitude in the zenith is equal to

$$h_B(\zeta=0, z) = (R+h_P)\sin^{-1} z - R \approx h_P + \frac{R\delta^2}{2}, \quad (4)$$

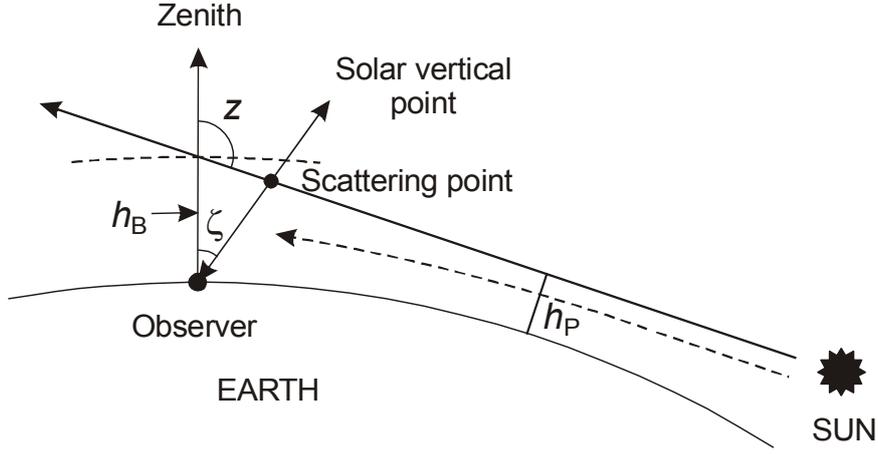

*Figure 2. Simple model of single scattering geometry during the twilight.*

Here $R$ is the radius of the Earth and $\delta = z - \pi/2$ is the depression of Sun below the horizon. We neglect the refraction again since it is noticeable only for lower rays transferring just above the shadow. In fact, decrease of scattering altitude of these rays is compensated by refraction divergence effect (dashed line in Figure 2) and extinction, especially if the aerosol particles are present in lower stratosphere.

We also note that the distance to the scattering point is significantly less than radius of the Earth, and the angle $\delta$ is small. Then the shadow altitude in any point of solar vertical can be written as

$$h_B(\zeta, z) = h_B(0, z) \cdot (1 - \delta \operatorname{tg}\zeta) = ((R+h_P)\sin^{-1} z - R) \cdot (1 - \delta \operatorname{tg}\zeta). \quad (5)$$

For the pressure values ratio in formula (3), we have

$$\frac{p(h_B(\zeta, z))}{p(h_B(0, z))} = \exp(-K(h_B(\zeta, z) - h_B(0, z))) = \exp(K((R+h_P)\sin^{-1} z - R)\delta \operatorname{tg}\zeta), \quad (6)$$

The coefficient $K$ is

$$K = \frac{\mu\, g(h_B(0, z))}{\Re T(h_B(0, z))}. \quad (7)$$

Here $\mu$ is molar mass of atmospheric substance, $g$ is the gravitational acceleration, $\Re$ is universal gaseous constant, and $T$ is the temperature.

Single molecular scattering is polarized, the value is

$$Q = \frac{\sin^2(z-\zeta)}{1.06 + \cos^2(z-\zeta)}. \quad (8)$$

We assume that single scattering above 40 km is principally molecular. It is added to the multiple scattering with intensity $j(\zeta, z)$ and polarization $q(\zeta, z)$. The last value is positive if the polarization direction is the same as for single scattering (perpendicular to the scattering plane, i.e. perpendicular to the solar vertical) and negative if the light is polarized along the scattering plane. Not assuming any intensity properties, we suppose that multiple scattering polarization is symmetric relatively the zenith:

$$q(\zeta, z) = q(-\zeta, z). \tag{9}$$

The total intensity and polarization of the sky background are equal to:

$$I(\zeta,z) = J(\zeta,z) + j(\zeta,z); \quad P(\zeta,z) = \frac{J(\zeta,z)Q(\zeta,z) + j(\zeta,z)q(\zeta,z)}{J(\zeta,z) + j(\zeta,z)}. \tag{10}$$

The values measured experimentally are ln $I$, $P$, and their derivatives by solar zenith angle $z$. For any definite $z$ value, we have to find the single scattering contribution in the zenith

$$S(z) = \frac{J(0,z)}{I(0,z)} \tag{11}$$

and the value of $K$ (formulae 6-7). We express the observational characteristics of multiple scattering in terms of $S$ and $K$. The intensity is

$$j(\zeta,z) = I(\zeta,z) - J(\zeta,z) = $$
$$= I(\zeta,z) - I(0,z)S(z)\frac{1.06 + \cos^2(z-\zeta)}{\cos\zeta \cdot (1.06 + \cos^2 z)}\exp(K((R+h_P)\sin^{-1}z - R)\delta \operatorname{tg}\zeta). \tag{12}$$

Polarization of multiple scattering is equal to

$$q(\zeta,z) = \frac{I(\zeta,z)P(\zeta,z) - J(\zeta,z)Q(\zeta,z)}{I(\zeta,z) - J(\zeta,z)}. \tag{13}$$

We repeat the same procedure for close twilight stage with solar zenith angle $z+dz$. The measured sky intensity and polarization values are:

$$I(\zeta, z+dz) = \exp(\ln I(\zeta,z) + dz\frac{d\ln I(\zeta,z)}{dz});$$
$$P(\zeta, z+dz) = P(\zeta,z) + dz\frac{dP(\zeta,z)}{dz}. \tag{14}$$

Derivatives in the right side of equation are also measured. Angle of single molecular scattering in the zenith is close to 90° (minimum of last term in formula (1)), and intensity changes due to shadow altitude evolution:

$$J(0,z) = I(0,z) \cdot S(z);$$
$$J(0,z+dz) = J(0,z)\frac{p(h_B(0,z+dz))}{p(h_B(0,z))} = J(0,z)\exp(-K(R+h_P)(\sin^{-1}(z+dz) - \sin^{-1}z)) \approx $$
$$\approx J(0,z)\exp(-K(R+h_P)\delta\, dz). \tag{15}$$

Using equations (3-7), we can find the intensity of molecular scattering in solar vertical $J(\zeta, z+dz)$ with substitution of required value of solar zenith angle, and also its polarization by equation (8). Finally, we find the multiple scattering polarization value by formula (13) and also its derivative by $z$:

$$q'(\zeta,z) = \frac{q(\zeta, z+dz) - q(\zeta,z)}{dz}. \tag{16}$$

Correction criterion of values $S$ and $K$ is the similarity of multiple scattering polarization and its derivative value in symmetrical solar vertical points:

$$\sum_{\zeta=0}^{\zeta_{MAX}} (q(\zeta,z) - q(-\zeta,z))^2 + Z_S^2(z) \sum_{\zeta=0}^{\zeta_{MAX}} (q'(\zeta,z) - q'(-\zeta,z))^2 = \min. \quad (17)$$

Here $Z_S$ is the scale of twilight sky brightness change in the zenith, interval of solar zenith angles corresponding to the decrease of intensity in $e$ times:

$$Z_S(z) = \left| \frac{d \ln I(0,z)}{dz} \right|^{-1}. \quad (18)$$

The value of $Z_S$ is close to 1° during the transitive twilight. So, we have to find two values $S$ and $K$ by nonlinear least-square procedure. The sense of checking of symmetry of derivative values $q'$ along with polarization $q$ itself is defined by additional requirements to the value of $K$ by equation (15). This value is related with Boltzmann temperature $T$ by formula (7). The values of $K$ and $T$ obtained in this procedure are model-dependent, we denote them as $K_1$ and $T_1$. Repeating the process for wide range of solar zenith angles or scattering altitudes, we find the dependence of sky brightness in the zenith on $z$:

$$J(0,z) = I(0,z) \cdot S(z) = \text{const} \cdot p(h_B(0,z)) \cdot (1.06 + \cos^2 z). \quad (19)$$

The amount of $\cos^2 z$ during the twilight stage being considered does not exceed 0.02, and the change of last term is incomparably less than the change of pressure by orders of value. Logarithmic derivative of single scattering brightness in the zenith $J(0, z)$ on the shadow altitude $h_B$ is equal to

$$\frac{d \ln J(0,z)}{dh_B(0,z)} = \frac{d \ln J(0,z)}{dz} \cdot \frac{dz}{dh_B(0,z)} = -K_2 = -\frac{\mu\, g(h_B(0,z))}{\Re T_2(h_B(0,z))}. \quad (20)$$

This relation gives the possibility of second independent estimation of temperature $T$. However, its accuracy is influenced by errors of values $J(0, z)$ contributed by definition of function $S(z)$ by least-square method (equations 11-17). Despite of it, the accuracy of $T_2$ value is better than for $T_1$ owing to sharp dependence of pressure by altitude. Final value of $T$ is found by averaging of $T_1$ and $T_2$ with weights back proportional to the squares of their errors.

Besides of primal estimations of temperature, the least-square procedure of minimization of asymmetry of multiple scattering (17) allows finding the contribution of multiple scattering in the twilight background and its polarization during the different stages of twilight, that is also interesting. The results of such analysis are listed in the next chapter of the paper.

4. RESULTS

At the first stage of the procedure described above, observational data is analyzed for each definite twilight moment. Figure 3 shows the profiles of intensity and polarization of molecular and multiple scattering and total background along the solar vertical at solar zenith angle 96° (effective shadow altitude in the zenith 50 km) in the evening twilight of May, 31, 2014. The dependency of molecular scattering contribution in total background is also plotted. Maximum of single scattering polarization is shifted to the dusk area, since it is placed at 90° from the Sun. The same maximum of total background is shifted by the less angle due to contribution of multiple scattering which polarization is symmetric relatively the zenith. This effect was used to estimate the multiple scattering contribution during the light twilight [19, 20].

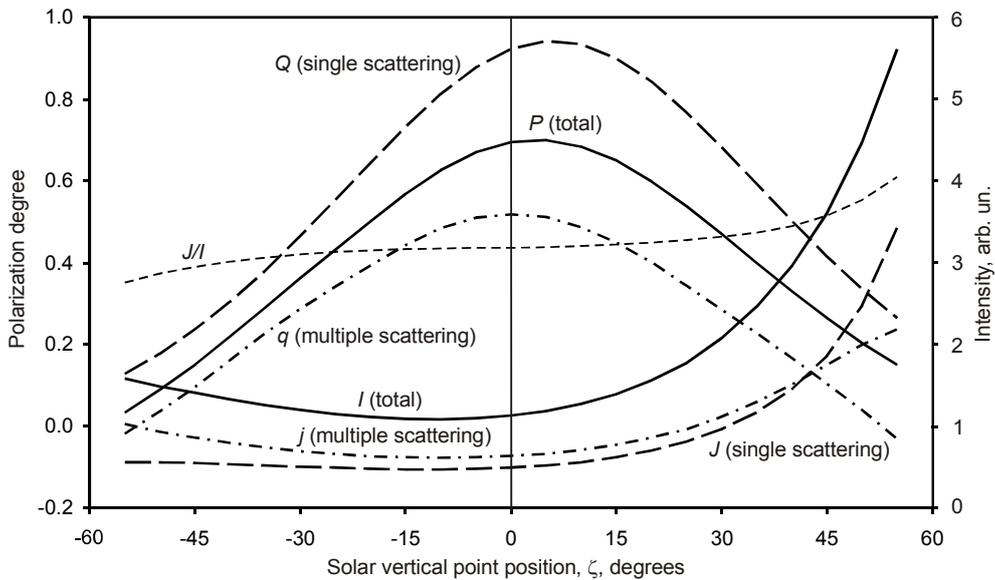

*Figure 3. Intensity and polarization of sky background and its components in solar vertical at solar zenith angle 96° (effective shadow altitude in the zenith 50 km), the same twilight as in Figure 1.*

Multiple scattering has the inverted polarization ($q<0$) at large point zenith angles. It is caused by scattering of side parts of the sky background [24]. Appearance of neutral points at the zenith angles 50°-60° and their motion during the twilight is related with change of the ratio of single and multiple scattering intensities.

At the same time, multiple scattering is significantly polarized near the zenith. This can be explained by scattering of glow of dusk segment. Owing to it, the total sky polarization remains high even during the dark twilight period, when single scattering fades away (see Figure 1). Analysis made in this work shows that multiple scattering polarization changes weakly during the transitive twilight, remaining almost the same as during the dark twilight (dashed lines in Figure 1). The same conclusion was made before [19, 28] and refers to all solar vertical points.

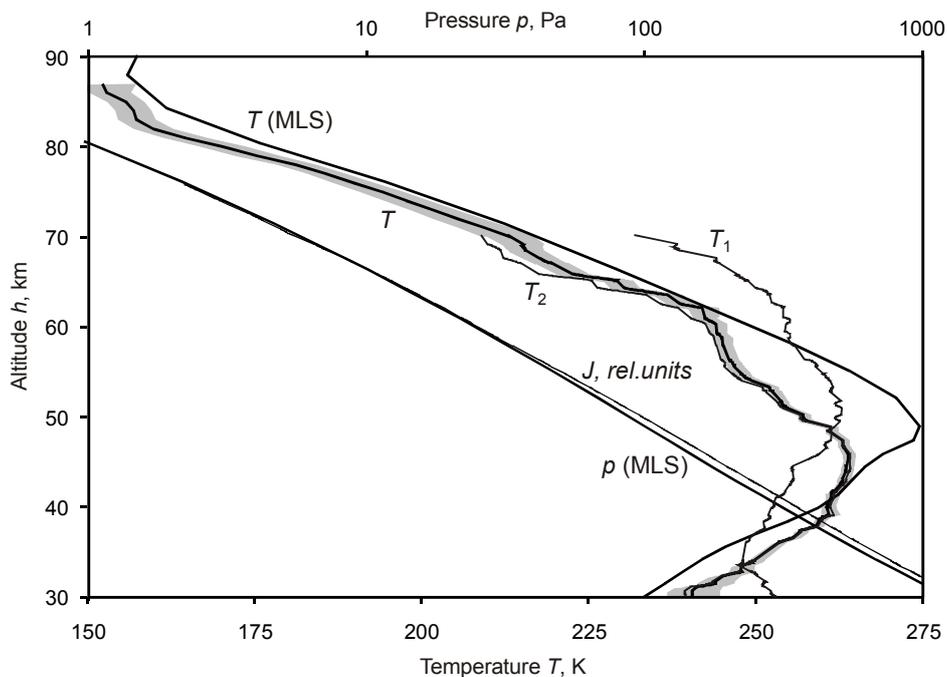

*Figure 4. Temperature profile in stratosphere and mesosphere compared with MLS data; single scattering profile compared with MLS pressure profile, the same twilight as in Figures 1 and 3.*

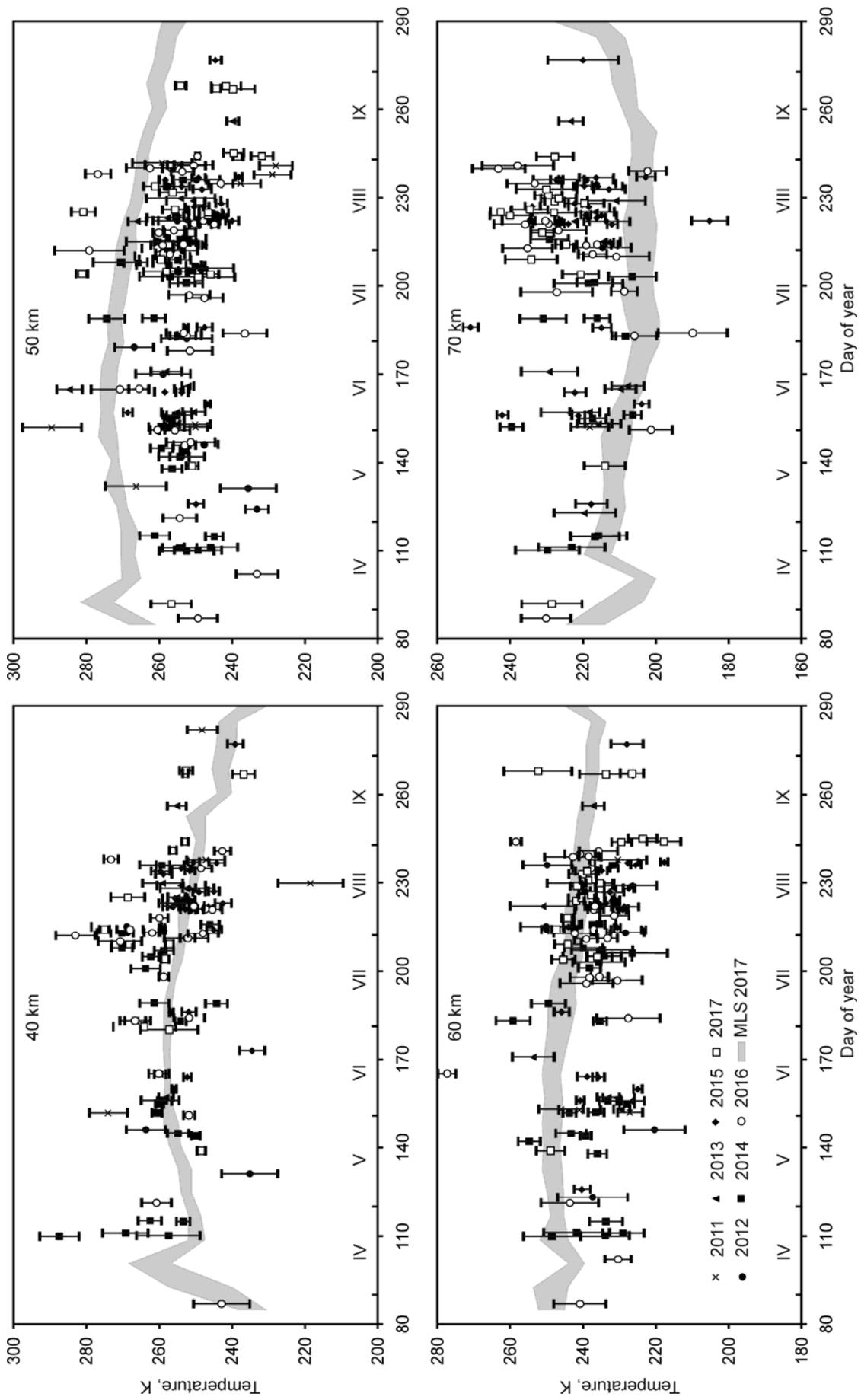

*Figure 5. Temperatures at 40-70 km based on sky background measurements in the vicinity of Moscow.*

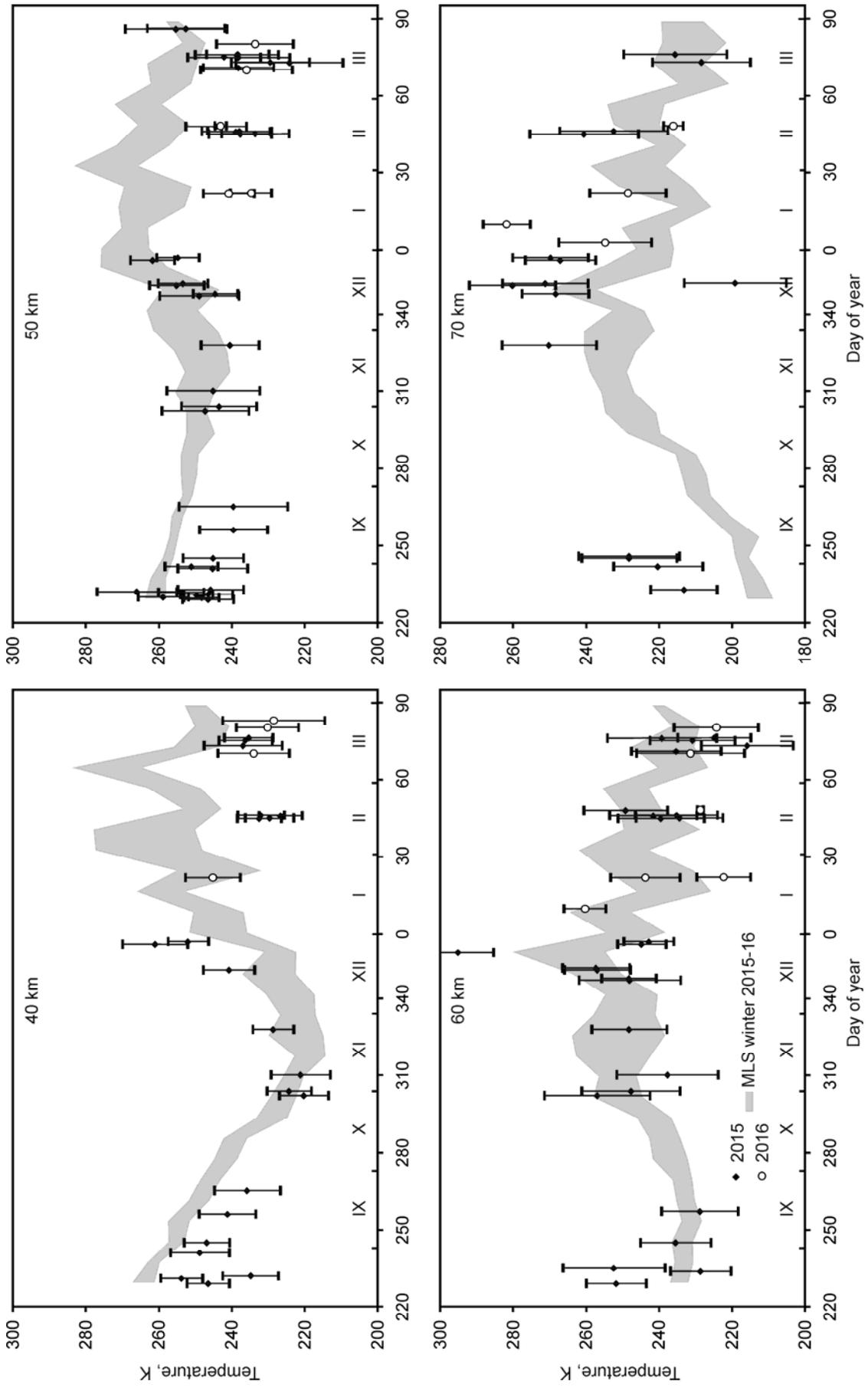

*Figure 6. Temperatures at 40-70 km based on sky background measurements in Apatity.*

Vertical profile of quantity $J(0, z)$ – the intensity of molecular scattering in the zenith depending on shadow altitude $h_B(0, z)$ for the same evening twilight of May, 31, 2014 (vicinity of Moscow) is shown in Figure 4. We also plot the pressure profile for the same day and nearby locations by EOS Aura/MLS data. We see that profiles are close to each other. This points to the satisfactory accuracy of separation method described above (note that intensity is expressed in arbitrary units). This allows estimating the temperature $T_2$ and averaged value $T$.

Results of this are also shown in Figure 4. In most part of cases the accuracy of temperature estimation $T_2$ is significantly better than for $T_1$ (their profiles are also shown), and the final value $T$ is close to $T_2$. Results are compared with EOS Aura/MLS profile. Twilight data clearly shows the stratopause temperature maximum near 45-50 km, however, it is blurred if we compare it with satellite profile. It can be related with restricted altitude resolution of twilight method: owing to extinction of tangent solar emission the shadow of the Earth has no sharp border. There are some other reasons that will be discussed below. This leads to underestimation of stratopause temperature by about 15K, the temperatures above and below are closer to satellite values. The twilight temperature profile above 70 km is built by the method described in [22]. Both procedures cover 50-km layer of the atmosphere.

Season dependencies of temperature at four levels measured in the vicinity of Moscow in 2011-2017 are plotted in Figure 5. The data points with accuracy better than 10K and difference between estimated values of $T_1$ and $T_2$ less than 20K are shown. 1σ-intervals of EOS Aura/MLS temperatures averaged over 8-day periods in 2017 are also plotted. Systematical differences of temperatures are seen at all altitudes except 40 km, especially for stratopause – 50 km. However, the annual changes of temperature with maximum at summer solstice are seen in the twilight data.

The same results for autumn-winter seasons in Apatity in 2015-2016 with accuracy better than 15K are shown in Figure 6. Systematical difference at stratopause is also seen, but principal details of temperature evolution including fast warming during the last days of 2015 are fixed basing on twilight measurements.

5. DISCUSSION AND CONCLUSION

In this paper we suggest the simple procedure of separation of single molecular and multiple scattering during the transitive twilight basing on polarization analysis performed by all-sky cameras. We use the elementary model of single scattering and make just one assumption about the properties of multiple scattering (polarization symmetry in solar vertical), that is confirmed by observations.

The main result of the work is the confirmation of principal possibility of separation of basic component of twilight glow in this model. This can be used, for example, for temperature retrieval. The profiles can have the systematical offsets at some altitudes including stratopause level. It can be related with low altitude resolution of twilight analysis and also with tiny aerosol particles in upper stratosphere. The light scattering properties of such particles can be close to Rayleigh law. This can be also confirmed by single scattering excess below 50 km compared with pressure profile (Figure 4). This shows the principal possibility to study upper stratospheric aerosol by twilight analysis. It is especially interesting since present satellite techniques (for example, OSIRIS (*Optical Spectrograph and InfraRed Imager System*) instrument onboard Odin satellite [34]) and color analysis [25] are restricted by altitudes 30-35 km. Obviously, this analysis requires more complicated atmospheric models to be used.

Another result of this work is high value of multiple scattering polarization near the zenith exceeding 0.5 during transitive twilight. Taking into account that total sky polarization is less than 0.8, the multiple scattering contribution is significant during the whole twilight period. Clarification of multiple scattering properties can expand the possibilities of twilight sounding method for analysis of molecular and aerosol scattering in the atmosphere.

**Acknowledgments.** The work is supported by Russian Foundation for Basic Research, grant 16-05-00170.

REFERENCES

1. Roble, R. G., and Dickinson, R. E., How will changes in carbon dioxide and methane modify the mean structure of the mesosphere and thermosphere? // *Geophysical Research Letters,* 1989, vol. 16, pp. 1441-1444.
2. Rind D., Shindell, D., Lonegran, P., Balachandran, N.K. Climate Change and the Middle Atmosphere. Part III: The Doubled CO2 Climate Revisited // *Journal of Climate*, 1998, vol. 11, pp. 876-894.
3. Ramaswamy, V., Chanin, M.-L., Angell, J., Barnett, J., Gaffen, D., Gelman, M., Keckhut, P., Koshelkov, Y., Labitzke, K., Lin, J.-J. R., O'Neill, A., Nash, J., Randel, W., Rood, R., Shine, K., Shiotani, M., Swinbank, R. Stratospheric temperature trends: Observations and model simulations // *Reviews of Geophysics,* 2001, vol. 39, pp. 71-122.
4. Thompson, D.W.J., Seidel, D.J., Randel, W.J., Zou, C.-Z., Butler, A.H., Mears, C., Osso, A., Long, C., Lin, R. The mystery of recent stratospheric temperature trends // *Nature*, 2012, vol. 491, pp. 692-697.
5. Golitsyn, G.S., Semenov, A.I., Shefov, N.N., Fishkova, L.M., Lysenko, E.V., and Perov, S.P., Long-term temperature trends in the middle and upper atmosphere // *Geophysical Research Letters*, 1996, vol. 23, pp. 1741-1744.
6. Beig, G., et al., Review of Mesospheric Temperature Trends // *Reviews of Geophysics,* 2003, vol. 41(4), pp. 1015-1055.
7. Vargin, P.N., Yushkov, V.A., Khaikin, S.M., Tsvetkova, N.D., Kostrykin, S.V., Volodin, E.M. Climate change and middle atmosphere – too many questions // *Vestnik Rossiiskoi Academii Nauk (Messenger of Russian Academy of Sciences)*, 2010, vol. 80, pp. 114-130 (in Russian).
8. Matsuno T. A dynamical model of the stratospheric sudden warming // *Journal of Atmospheric Science*, 1971, vol. 28, pp. 1479-1494.
9. Labitzke, K., & Naujokat, B. The lower Arctic stratosphere in winter since 1952 // *SPARC Newsletters*, 2000, vol. 15, pp. 11-14.
10. Nikolashkin, S.V., Titov, S.V., Marichev, V.N., Bychkov, V.V., Kurkin, V.I., Chernigovskaya, M.A., Nepomnyashii, Yu. A. Lidar study of winter sudden stratospheric warmings behavior at the territory of Siberia and Far East // *Nauka i Obrazovanie (Science and Education)*, 2013, vol. 69, pp. 10-17 (in Russian).
11. von Zahn, U., Fiedler, J., Naujokat, B., Langematz, U., Kruger, K. A note on record-high temperatures at the northern polar stratopause in winter 1997/98 // *Geophysical Research Letters*, 1998, vol. 25, pp. 4169-4172.
12. France, J.A., Harvey, V.L., Randall, C.E., Hitchman, M. H., Schwartz, M. J. A climatology of stratopause temperature and height in the polar vortex and anticyclones // *Journal of Geophysical Research,* 2012, vol. 117, p. D06116.
13. Zuev, V.V., Marichev, V.N., Bondarenko, S.L. Study of accuracy characteristics of temperature profiles retrieval based on lidar signals of molecular scattering // *Atmospheric and Oceanic Optics*, 1996, vol. 9, № 12, pp. 1615-1619.
14. Marichev, V.N., Bochkovsky, D.A. Lidar measurements of air density in middle atmosphere. Part 1. Modeling of potential possibilities in visible part of spectrum // *Atmospheric and Oceanic Optics*, 2013, vol. 26, № 7, pp. 553-563.

15. Russell, J.M. III, Mlynczak, M.G., Gordley, L.L., Tansock, J., and Esplin, R., An overview of the SABER experiment and preliminary calibration results // *Proc. SPIE Int. Soc. Opt. Eng.,* 1999, vol. 3756, pp. 277-288.
16. Schwartz, M.J., et al., Validation of the Aura Microwave Limb Sounder temperature and geopotential height measurements // *Journal of Geophysical Research*, 2008, vol. 113, p. D15S11.
17. Hauchecorne, A., & Chanin, M.-L., Density and temperature profiles obtained by lidar between 35 and 70 km // *Geophysical Research Letters,* 1980, vol. 7, pp. 565-568.
18. Rosenberg, G.V. Twilight. Plenum Press, New York. 1966.
19. Ugolnikov, O.S. Twilight sky photometry and polarimetry. The problem of multiple scattering at the twilight time // *Cosmic Research,* 1999, vol. 37, № 2, pp. 159-166.
20. Ugolnikov, O.S., Maslov I.A. Multi-color polarimetry of the twilight sky. The role of multiple scattering as the function of wavelength // *Cosmic Research*, 2002, vol. 40, № 3, pp. 224-232.
21. Ugolnikov O.S., Maslov I.A. Optical properties of the undisturbed mesosphere from wide-angle twilight sky polarimetry // *Cosmic Research,* 2013, vol. 51, № 4, pp. 235-240.
22. Ugolnikov, O.S., & Maslov, I.A., Summer mesosphere temperature distribution from wide-angle polarization measurements of the twilight sky // *Journal of Atmospheric and Solar-Terrestrial Physics*, 2013, vol. 105-106, pp. 8-14.
23. Ugolnikov, O.S., Kozelov, B.V. Study of the mesosphere using wide-field twilight polarization measurements: early results beyond the polar circle // *Cosmic Research*, 2016, vol. 54, № 4, pp. 279-284.
24. Ugolnikov, O.S., Maslov, I.A. Analysis of the direction of the twilight sky background polarization as a tool for selecting single scattering // *Cosmic Research*, 2017, vol. 55, № 3, pp. 169-177.
25. Ugolnikov, O.S., Maslov, I.A. Background stratospheric aerosol investigations using multi-color wide-field measurements of the twilight sky // *Cosmic Research*, 2018, vol. 56, № 2.
26. Ugolnikov O.S., Maslov I.A., Kozelov B.V., Dlugach J.M. Noctilucent cloud polarimetry: Twilight measurements in a wide range of scattering angles // *Planetary and Space Science*, 2016, vol. 125, pp.105-113.
27. Ugolnikov O.S., Galkin A.A., Pilgaev S.V., Roldugin A.V. Noctilucent Cloud Particle Size Determination based on Multi-Wavelength All-Sky Analysis // *Planetary and Space Science*, 2017, vol. 146, pp. 10–19.
28. Ugolnikov O.S., Maslov I.A. Detection of Leonids meteoric dust in the upper atmosphere by polarization measurements of the twilight sky // *Planetary and Space Science*, 2007, vol. 55, pp.1456-1463.
29. Ugolnikov O.S., Maslov I.A. Studies of the stratosphere aerosol layer based on polarization measurements of the twilight sky // *Cosmic Research*, 2009, vol. 47, № 3, pp. 198-207.
30. Ugolnikov O.S., Maslov I.A. Polarization studies of contribution of aerosol scattering to the glow of twilight sky // *Cosmic Research*, 2005, vol. 43, № 6, pp. 404-412.
31. Postylyakov O.V. Linearized vector radiative transfer model MCC++ for a spherical atmosphere. // *Journal of Quantitative Spectroscopy and Radiative Transfer*, 2004, vol. 88, pp. 297-303.
32. Ugolnikov O.S., Postylyakov O.V., Maslov I.A. Effects of multiple scattering and atmospheric aerosol on the polarization of the twilight sky // *Journal of Quantitative Spectroscopy and Radiative Transfer*, 2004, vol. 88, pp. 233-241.
33. Zaginailo Yu.I. Determination of the second twilight brightness by the method of the twilight probing of the Earth's atmosphere. // *Odessa Astronomical Publication*, 1993, vol. 6, pp. 59-67.
34. Bourassa, A.E., Degenstein, D.A., Llewellyn, E.J. Retrieval of stratospheric aerosol size information from OSIRIS limb scattered sunlight spectra // *Atmospheric Chemistry and Physics Discussions*, 2008, vol. 8, pp. 4001-4016.